\documentclass[a4paper,twoside]{article}

\usepackage{epsfig}
\usepackage{subcaption}
\usepackage{calc}
\usepackage{amssymb}
\usepackage{amstext}
\usepackage{amsmath}
\usepackage{amsthm}
\usepackage{multicol}
\usepackage{pslatex}
\usepackage{apalike}
\usepackage{algorithm2e}
\usepackage[bottom]{footmisc}
\usepackage{url}                               
\usepackage[hidelinks,dvipdfmx]{hyperref}      
\usepackage{multirow}                          
\usepackage{graphicx}                          
\usepackage[dvipdfmx]{color}                   
\usepackage{SCITEPRESS}     
\begin{document}

\title{Towards Interoperable Data Spaces: Comparative Analysis of Data Space Implementations between Japan and Europe}

\author{\authorname{Shun Ishihara\sup{1}\orcidAuthor{0009-0005-4584-2695}, Taka Matsutsuka\sup{1}\orcidAuthor{0000-0002-2673-6708}}
\affiliation{\sup{1}Fujitsu Limited, Kawasaki, Kanagawa 211-8588, Japan}
\email{\{ishihara.shun, markn\}@fujitsu.com}
}


\abstract{Data spaces are evolving rapidly. In Europe, the concept of data spaces, which emphasises the importance of trust, sovereignty, and interoperability, is being implemented as a platform such as Catena-X. Meanwhile, Japan has been developing its approach to data sharing, in line with global trends but also to address unique domestic challenges, resulting a platform such as DATA-EX. Achieving interoperability between European and Japanese data spaces remains a critical challenge due to the differences created by these parallel advances. Although interoperability between data spaces has several aspects, compatibility of trust in the participating entities and the data exchanged is a significant aspect due to its influence on business. This paper undertakes a comparative analysis of DATA-EX and Catena-X while focusing on aspect of trust, to explore the challenges and opportunities for achieving interoperability between Japanese and European data spaces. By examining common data exchange processes, key objects such as datasets, and specific evaluation criteria, the study identifies gaps, challenges, and proposes actionable solutions such as inter-exchangeable topology. Through this analysis, the paper aims to contribute to the ongoing discourse on global data interoperability.}

\onecolumn \maketitle \normalsize \setcounter{footnote}{0} \vfill

\section{\uppercase{Introduction}}
\label{sec:introduction}

The rapid evolution of data spaces is transforming the landscape of secure and interoperable data sharing across industries and regions. In Europe, the concept of data spaces, supported by initiatives such as the European Data Strategy, emphasises the importance of trust, sovereignty, and interoperability. Meanwhile, Japan has been developing its approach to data sharing, in line with global trends but also addressing unique domestic challenges. Despite these parallel advancements, achieving interoperability between European and Japanese data spaces remains a critical challenge due to differences in governance, technology standards, and authentication frameworks. Interoperability between data spaces has various aspects such as compatibility of technologies like connectors, compatibility of objects handled in data spaces, and compatibility of trust on participating entities and the data exchanged. Compatibility of trust is a significant issue among these aspects, because actual data exchange is never established without trustworthiness between participants.

This paper undertakes a comparative analysis of DATA-EX and Catena-X to explore the challenges and opportunities for achieving interoperability between Japanese and European data spaces. The analysis is focusing on trust. By examining common data exchange processes, key objects such as participants, datasets, and data catalogs, and specific criteria for evaluation, the study identifies gaps and proposes actionable solutions. The findings are based on a thorough analysis of relevant documentation, highlighting areas for alignment and improvement.

Through this analysis, the paper aims to contribute to the ongoing discourse on global data interoperability. It proposes an inter-exchangeable topology that bridges regional differences while addressing common challenges.

The rest of the paper is structured as follows. Section \ref{sec:background} reviews the development of data spaces in Europe and Japan, and examines the role of trust anchors and interoperability challenges. Section \ref{sec:analysismethod} describes the analysis method, including key evaluation criteria and data sources. Section \ref{sec:analysisresult} presents the analysis results, comparing participants, devices, datasets, and other key elements of the two platforms. Section \ref{sec:discussion} discusses identified gaps, proposes an inter-exchangeable topology, and addresses specific challenges. Section \ref{sec:relatedworks} reviews related works, and Section \ref{sec:conclusion} concludes with findings and future directions.

\section{\uppercase{Background}}
\label{sec:background}

In recent years, the importance of the data economy has been steadily increasing, with data spaces evolving through unique approaches in different regions and countries. Data spaces refer to infrastructures or frameworks that enable the sharing and integration of data, allowing participating organisations and individuals to maintain data sovereignty while securely sharing information. This section provides an overview of the development of data spaces in Europe and Japan, highlights their characteristics, and identifies challenges for the future.

\subsection{Development of Data Spaces in Europe}

Europe has been accelerating its efforts to maximise the potential of the data economy. Central to these efforts are the European Data Strategy\footnote{\url{https://commission.europa.eu/strategy-and-policy/priorities-2019-2024/europe-fit-digital-age/european-data-strategy_en}} and the concept of data sovereignty. The European Data Strategy aims to foster data sharing within and beyond the region, positioning Europe as a global leader in the data economy. This strategy is rooted in the belief that data should be managed and utilised fairly, without being disproportionately controlled by a few companies or nations. Data sovereignty refers to the principle that entities owning and managing data have the exclusive right to determine its usage. For example, this includes the control over the data, even if the data is shared with other parties. Europe has reinforced this principle through stringent data protection regulations such as General Data Protection Regulation (GDPR), ensuring that individuals and organisations can safeguard their rights while contributing to the data economy. Initiatives like International Data Spaces Association (IDSA) and Gaia-X exemplify Europe's approach to data spaces. IDSA focuses on creating secure and standardised frameworks for data sharing across various industries, ensuring that data sovereignty is maintained. Meanwhile, Gaia-X, a European initiative for cloud infrastructure and data sharing, aims to establish an interoperable ecosystem where data and services can be securely exchanged while adhering to European values such as transparency and privacy. Together, these initiatives highlight Europe's commitment to fostering a trusted data economy by addressing technical, legal, and ethical challenges inherent in cross-border data exchanges. With this multipronged strategy, Europe seeks to establish itself as a leader in the development of a robust, scalable, and ethical data ecosystem.

\subsection{Development of Data Spaces in Japan}

Similarly, Japan has pursued its own approach to developing data spaces, centerd on the concepts of Society 5.0\footnote{\url{https://www8.cao.go.jp/cstp/english/society5_0/}} and Data Free Flow with Trust (DFFT)\footnote{\url{https://www.digital.go.jp/en/policies/dfft}}. Society 5.0 is a national strategy that envisions a super-smart society where cyberspace and physical space are seamlessly integrated to address societal challenges through data utilisation. Within this concept, DFFT emphasises the importance of trust in data flow, enabling secure data sharing across national and organisational boundaries. The second phase of the Strategic Innovation Promotion Program (SIP) and its subsequent evolution into DATA-EX \cite{10020855} have constituted significant advances in Japan's data space endeavours, spearheaded by governmental initiatives. In the context of SIP, a connector called Connector Architecture for Decentralized Data Exchange (CADDE)\footnote{\url{https://sip-cyber-x.jp/en/}} has been developed with the objective of enabling secure and efficient data sharing between organisations. This foundational technology constituted the basis for the subsequent establishment of the DATA-EX initiative, which was launched within Data Society Alliance (DSA). The objective of this initiative is to expand the capabilities of CADDE into a broader data exchange infrastructure, thereby promoting interoperability and trust in data sharing across sectors. Another notable effort is the Ouranos Ecosystem \cite{OURANOS} project, which began in 2022 under the leadership of Information Processing Agency's Digital Architecture Design Center (IPA DADC). The project's objective is to enhance battery traceability within the automotive manufacturing supply chain as well as ensure compliance with European battery regulations. This addresses the critical need for transparency and accountability in tracking battery production and usage. In 2023, the implementation phase was completed, establishing a technical foundation for real-world application. By 2024, the operational body, Automotive and Battery Traceability Center Association (ABtC), was established, marking the transition to full-scale operation and contributing to the advancement of sustainable supply chain management.

\subsection{The Importance of Trust Anchors in Data Spaces}

Trust anchors are critical for ensuring the security, authenticity, and reliability of data spaces, enabling participants to confidently share data while maintaining sovereignty. Key technologies include digital identity systems like decentralised identifiers (DIDs), which allow entities to manage their own digital identities securely, and verifiable credentials (VCs), which provide proof of attributes or certifications without exposing unnecessary information. Governance frameworks, like those from IDSA, establish common rules for interoperability and trust. Together, these technologies and standards build a secure and reliable foundation, fostering participation and unlocking the full potential of global data spaces.

\subsection{Challenges of Interoperability Between Japan and Europe}

Currently, data spaces in Japan and Europe are built on different technological bases, making interoperability challenging. In Europe, standardisation of data space and trust is being pursued in IDSA and Gaia-X respectively. In Japan, meanwhile, the DATA-EX and Ouranos Ecosystem are pursuing their own individual standardisation agendas. However, the need for interoperability between these data spaces is becoming increasingly critical, particularly for addressing global issues such as carbon footprint transparency and international supply chain management. Interoperability between data spaces has various aspects such as compatibility of technologies which means connectors or related components can connect each other seamlessly beyond a data space, compatibility of objects which means expression of objects handled in data spaces can be interpreted uniquely, and compatibility of trust which means trustworthiness of participating entities and the data exchanged is similar among data spaces. A significant challenge lies in the compatibility of trust. Without robust mechanisms to guarantee trust, resistance to data sharing may increase, stalling the flow of data and hindering the data economy. Therefore, creating systems that integrate or harmonise the technological foundations of Japan and Europe is essential. Addressing this challenge will pave the way for a sustainable and interconnected data economy in the future.

\section{\uppercase{ANALYSIS METHOD}}
\label{sec:analysismethod}

As a first step towards solving the challenge mentioned above, we undertake a comparative analysis of two prominent data space platforms in Japan and Europe from a trust perspective. We selected DATA-EX in Japan and Catena-X in Europe as the representative data space platforms due to substantial number of analysable documents and the past maturity evaluation \cite{dam2023survey} \cite{bacco2024data}.

In this section, we examine a common data exchange process, key objects such as participants, datasets, and data catalogs, and specific criteria for evaluation, and available documents used for analysis.

\subsection{Common Data Exchange Process}

First, we decided on a common data exchange process to make a fair comparison between the two platforms. We researched principal whitepapers written in Japan and Europe to compare regional differences. 

A whitepaper from Data Society Alliance \cite{DSA} described processes, functions and elements when data providers and data consumers who were in different organisations or departments plan to handle data across existing systems or applications. The paper attempted to define the data exchange process. Six phases below were extracted as the data exchange process in the paper.

\begin{itemize}
  \item Planning: A data provider prepares the data to be transferred. A data consumer designs a new service or application to use the data and defines data requirements.
  \item Discovery: The data provider describes and publishes the catalog of the data. The data consumer searches for the data to meet requirements through a catalog search service.
  \item Contract: The data provider and the data consumer negotiate terms and conditions of the data exchange and conclude a contract.
  \item Transfer: The data provider makes the data available in an accessible location. The data consumer obtains the data from the provider and uses it within scope of the contract.
  \item Payment: The data consumer pays for the data based on the contract.
  \item Verification: The data provider and the data consumer verify the status of the contract fulfilment (e.g., progress of data delivery or payment).
\end{itemize}

A whitepaper from International Data Spaces Association \cite{IDSRAM4} defined a reference architecture model to generalise concepts, functionality, and overall processes for creating a secure network of trusted data. Their reference architecture model defined five layers: business layer, functional layer, information layer, process layer, and system layer. The process layer defined five phases below as the interactions that take place between the different components (e.g., connectors) of the international data spaces.

\begin{quotation}
  1. Onboarding, i.e. what to do to be granted access to the International Data Spaces as a Data Provider or Data Consumer.

  2. Data Offering, i.e. offering data or searching for a suitable data.

  3. Contract Negotiation, i.e. accept data offers by negotiating the usage policies.

  4. Exchanging Data, i.e. transfer data between IDS Participants.

  5. Publishing and using Data Apps, i.e. interacting with an IDS App Store or using IDS Data Apps.
\end{quotation}

There are some overlaps between the phases of DSA and IDSA. Table~\ref{table1} shows the correspondence between DSA and IDSA phases.

\begin{table}[htbp]
  \caption{Correspondence between DSA and IDSA phases.}
  \centering
  \begin{tabular}{l|l} \hline
      DSA & IDSA \\ \hline\hline
      - & Onboarding \\ \hline
      Planning & - \\ \hline
      Discovery & Data Offering \\ \hline
      Contract & Contract Negotiation \\ \hline
      Transfer & Exchanging Data \\
      & Publishing and using Data Apps \\ \hline
      Payment & - \\ \hline
      Verification & - \\ \hline
  \end{tabular}
  \label{table1}
\end{table}

Considering this correspondence, we extract the onboarding phase from IDSA's process and all six phases from the DSA's process as a common data exchange process.

\subsection{Target Objects}

Based on the common data exchange process, we identified objects which are generated in a data exchange as a target for our analysis. (The term ``object'' here means a digital entity that requires some kind of trust.) As a result, six objects shown in Table~\ref{table2} were identified.

\begin{table}[htbp]
  \caption{Generated objects in the common data exchange process.}
  \centering
  \begin{tabular}{l|p{5cm}} \hline
      Phase & Generated object \\ \hline\hline
      Onboarding & Participant (data consumer or provider account) \\
      & Device (server and endpoint on which connector works) \\ \hline
      Planning & Dataset \\
      & Data catalog \\ \hline
      Discovery & - (only refer data catalogs) \\ \hline
      Contract & Contract \\ \hline
      Transfer &  Sending \& Receiving Log \\
      Payment & \\ \hline
      Verification & - (only refer previously generated log) \\ \hline
  \end{tabular}
  \label{table2}
\end{table}

In the onboarding phase, a data space operating company validates an applicant organisation and then creates a user account. We defined this account as a ``Participant'' object. Following the acquisition of the user account, the applicant organisation prepares servers and endpoints on which the connectors operate. We defined these servers and endpoints as ``Device'' objects. During the planning phase, a data provider prepares datasets to provide and data catalogs to advertise their datasets. We defined these two as ``Dataset'' and ``Data catalog'' objects. In the contract phase, the data provider and consumer conclude a contract about a data exchange. A record of the contract is stored as a paper or a digital data. We defined these records as ``Contract'' objects. In the transfer and payment phases, actual data and money are transferred in accordance with the contract. The records of these transactions are stored as a digital log to prove fulfilment of the contract. We defined these logs as ``Sending \& Receiving Log'' objects.

\subsection{Analysis Criteria}

We have developed an analysis criteria on trust based on six analysis target objects.

In order for an object to be considered trustworthy, trust needs to be satisfied at two steps: when generated and when verified. The first step is to ensure trust of the object when it is generated. In the case of the participant object, an applicant is validated by a system administrator in a certain aspect (e.g., existence, eligibility, agreement to rules). The second step is to verify whether trust is ensured during the operation of the data space. In the case of the participant object, after the validation, the system administrator issues a credential (e.g., user id and password), and the system verifies the credential on login. Considering these two steps, we created two analysis perspectives.

\begin{itemize}
  \item p1) How to ensure trust in the object generation 
  \item p2) How to verify trust in operation
\end{itemize}

The final analysis criteria using identified objects and analysis perspectives is shown in Table~\ref{table3}.

\begin{table}[htbp]
  \caption{Analysis criteria.}
  \centering
  \begin{tabular}{l||l|l} \hline
      & p1 & p2 \\ \hline\hline
      Participant &  & \\ \hline
      Device &  & \\ \hline
      \dots &  & \\ \hline
  \end{tabular}
  \label{table3}
\end{table}

The horizontal axis represents analysis perspectives, while the vertical axis represents identified objects. Employing these criteria facilitates comprehensive analysis.

\subsection{Documents Used for Analysis}

Finally, we selected the technical documents to be used in the analysis.

With regard to DATA-EX, we selected technical documents in both the industrial data linkage infrastructure\footnote{\url{https://www.digital.go.jp/en/policies/industrial-data-integration}} and CADDE documents\footnote{\url{https://github.com/CADDE-sip}}.  Digital Agency is facilitating its implementation as industrial data linkage infrastructure and publishing its basic design documents. Note that some documents are not included in the industrial data linkage infrastructure's publication but in the CADDE publication because industrial data linkage infrastructure is an enhancement of CADDE.

With regard to Catena-X, we selected technical documents in both the technical standards \cite{CXSTD} and deliverables of the Eclipse Tractus-X project\footnote{\url{https://github.com/eclipse-tractusx}}. Catena-X Automotive Network e.V. is publishing technical standards, but it doesn't cover all specifications of Catena-X. Eclipse Tractus-X is an open-source project to develop reference implementations of Catena-X and providing some technical documents (not comprehensive) and source code. The version of the selected documents was adhered to CX-Jupiter release as Catena-X is under continuous development.

\section{\uppercase{ANALYSIS RESULT}}
\label{sec:analysisresult}

Summary of the analysis result is shown in Table~\ref{table4}.

\begin{table*}[htbp]
  \caption{Summary of the analysis result.}
  \centering
  \begin{tabular}{p{1.5cm}||p{2.8cm}|p{2.8cm}|p{2.8cm}|p{2.8cm}} \hline
      & \multicolumn{2}{|l|}{p1) How to ensure trust in the object generation} & \multicolumn{2}{|l}{p2) How to verify trust in operation} \\ \cline{2-5}
      & DATA-EX & Catena-X & DATA-EX & Catena-X \\ \hline\hline
      \multirow{2}{*}{Participant} & \multicolumn{2}{|l|}{Onboarding Portal} & \multirow{2}{2.8cm}{JWT (issued via OIDC protocol \& presented via original protocol)} & \multirow{2}{2.8cm}{VC (issued \& presented via original protocol)} \\ \cline{2-3}
      & Validation with DATA-EX specific standards & Validation with Catena-X specific standards & & \\ \hline
      \multirow{2}{*}{Device} & Validation with DATA-EX specific standards & Validation with Catena-X specific standards & \multicolumn{2}{l}{\multirow{2}{5.6cm}{Connector endpoint resolver and general HTTPS verification based on X.509 certificate}} \\ \cline{2-3}
      & \multicolumn{2}{|p{5.6cm}|}{Validation with CA's common standards (e.g., Extended Validation)} \\ \hline
      Dataset & \multicolumn{2}{|l|}{Data model \& vocabulary repository} & Signable data package & N/A \\ \hline
      Data catalog & \multicolumn{2}{|p{5.6cm}|}{Modified DCAT v2 format catalog creation function} & \multicolumn{2}{|l}{N/A} \\ \hline
      Contract & N/A & Connector’s contract negotiation API & \multicolumn{2}{|p{5.6cm}}{N/A (External contract brokering service is recommended, and connector can work with it)} \\ \hline
      Sending \& Receiving Log & Centralized log management service & N/A (local log) & Centralized log management service & N/A (local log) \\ \hline
  \end{tabular}
  \label{table4}
\end{table*}

In this summary, function or method of each data space is described. N/A means that there is no function or method. There is N/A in the dataset, data catalog, and contract, because the trust of these objects should be ensured by the participant in principle. Therefore, described function or method of these objects is not comprehensive solution but solution to complement the trust ensured by the participant. Details of the analysis results for each object are shown below.

\subsection{Participant}

\begin{figure}[htbp]
\centerline{\includegraphics[scale=0.3]{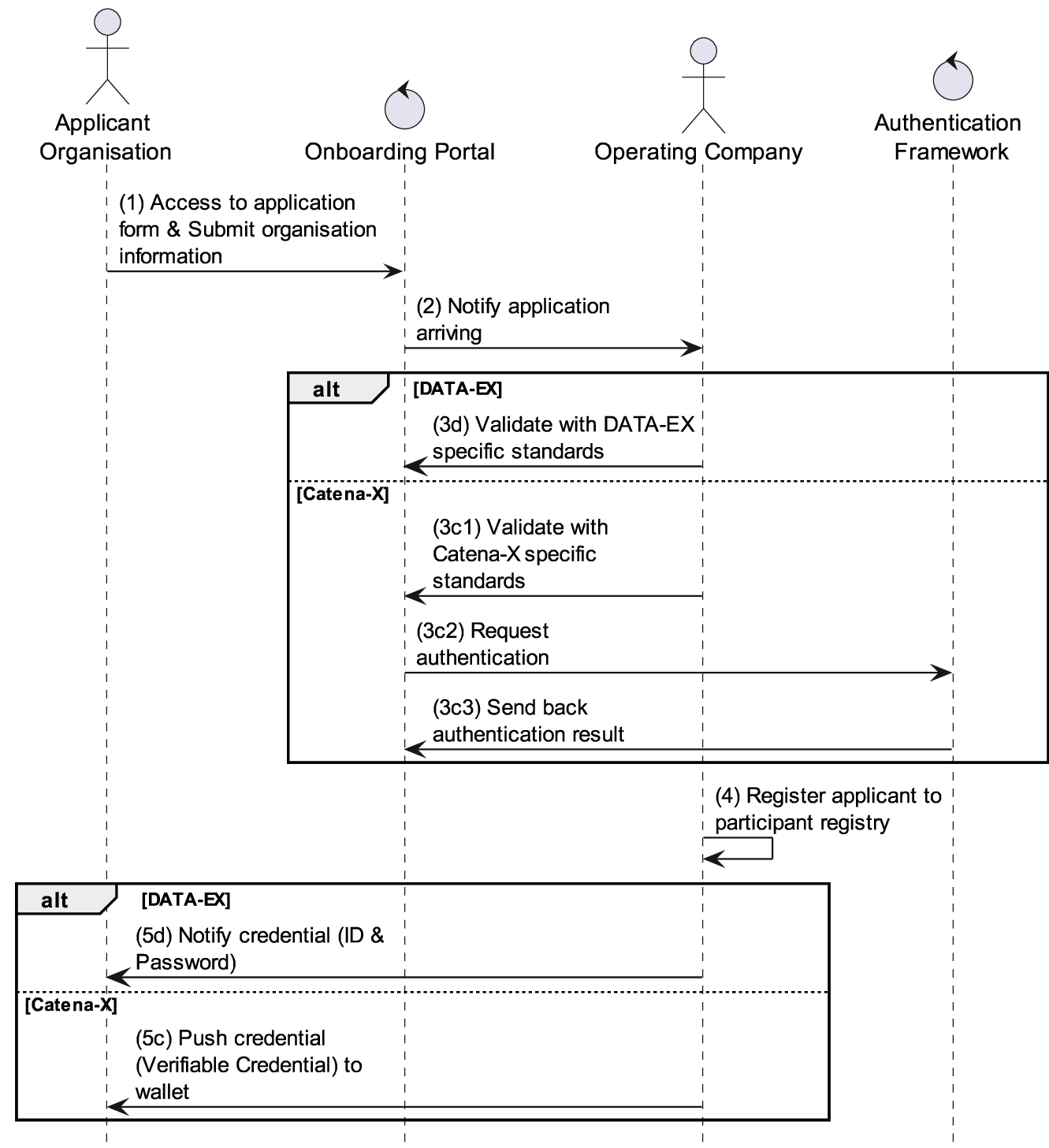}}
\caption{Validation process of the participant.}
\label{fig1}
\end{figure}

About p1, both platforms validate the applicant organisation through an onboarding portal. A validation process of each platform is shown in Figure~\ref{fig1}.

In both platforms, the applicant organisation submits organization information to the operating company through onboarding portal, but there are two differences.

The first difference is the validation method and standard (illustrated in Figure~\ref{fig1} as 3d,3c1,3c2,3c3). The operating company in both platforms validates similarly submitted information, but the validation standards are not the same. In addition, participant authentication is performed by an authentication framework in Catena-X. This framework is called as Gaia-X Digital Clearing House \cite{GXDCH}, which automatically validates an organisation from different perspectives (e.g., notarisation check with LEI code). This framework can be used commonly by multiple data spaces.  By using this framework, one data space can obtain compatibility about trust of the participant with another data space.

The second difference is a credential which the applicant organisation receives (illustrated in Figure~\ref{fig1} as 5d,5c). In DATA-EX, a participant registry is an identity provider, or IdP, so the applicant organisation is registered in form of an account and receives the id and password of the account as a credential. In Catena-X, on the other hand, the participant registry is a verifiable data registry (VDR), so the applicant organisation is registered in form of a DID document which holds a public key and receives a verifiable credential signed by a private key of the operating company.

\begin{figure}[htbp]
\centerline{\includegraphics[scale=0.3]{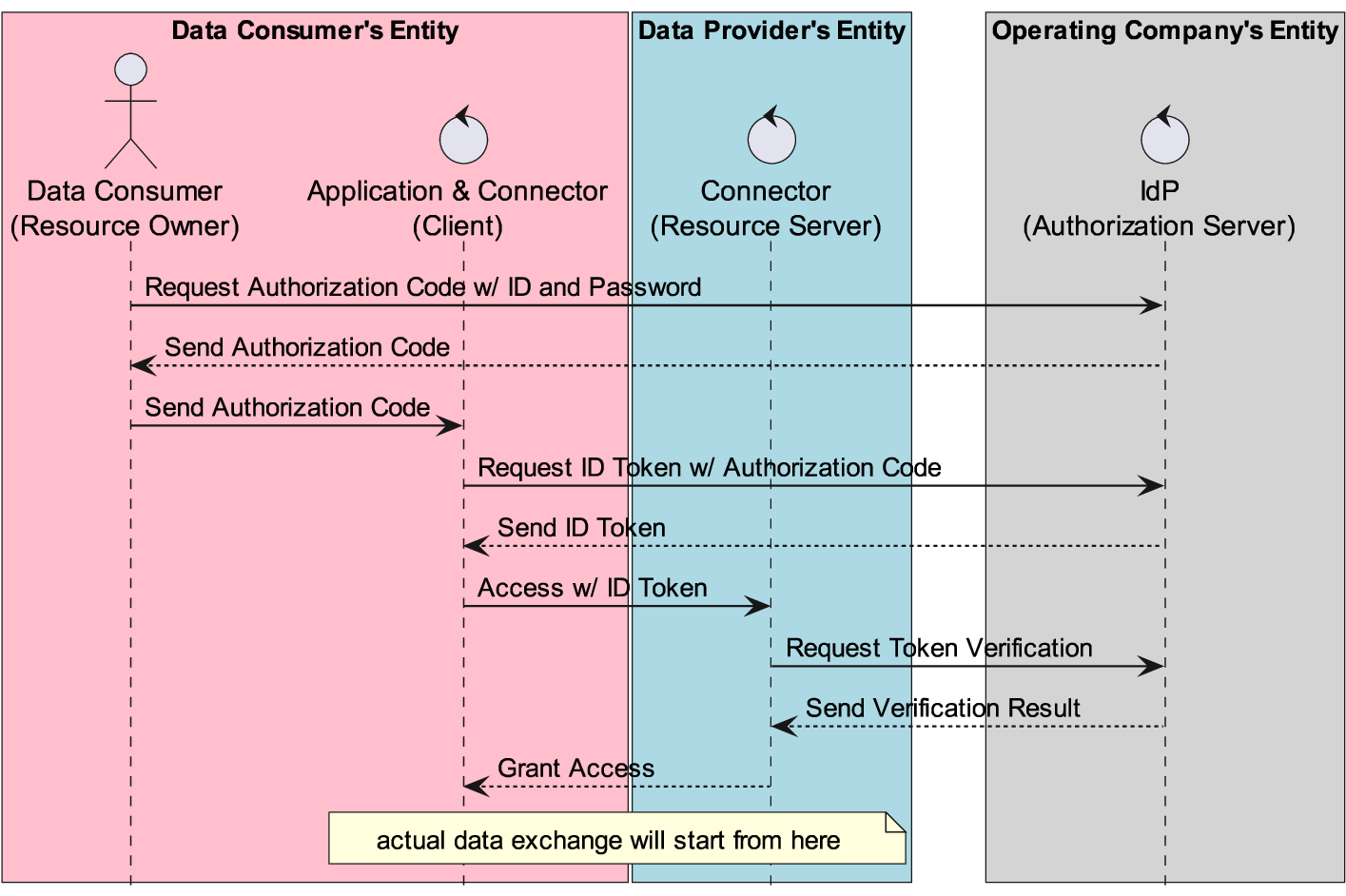}}
\caption{Verification process of the participant in DATA-EX.}
\label{fig2}
\end{figure}

The difference in the credential affects function or method on p2. In DATA-EX, the credential is handled via the OpenID Connect (OIDC) protocol shown in Figure~\ref{fig2}.

\begin{figure}[htbp]
\centerline{\includegraphics[scale=0.3]{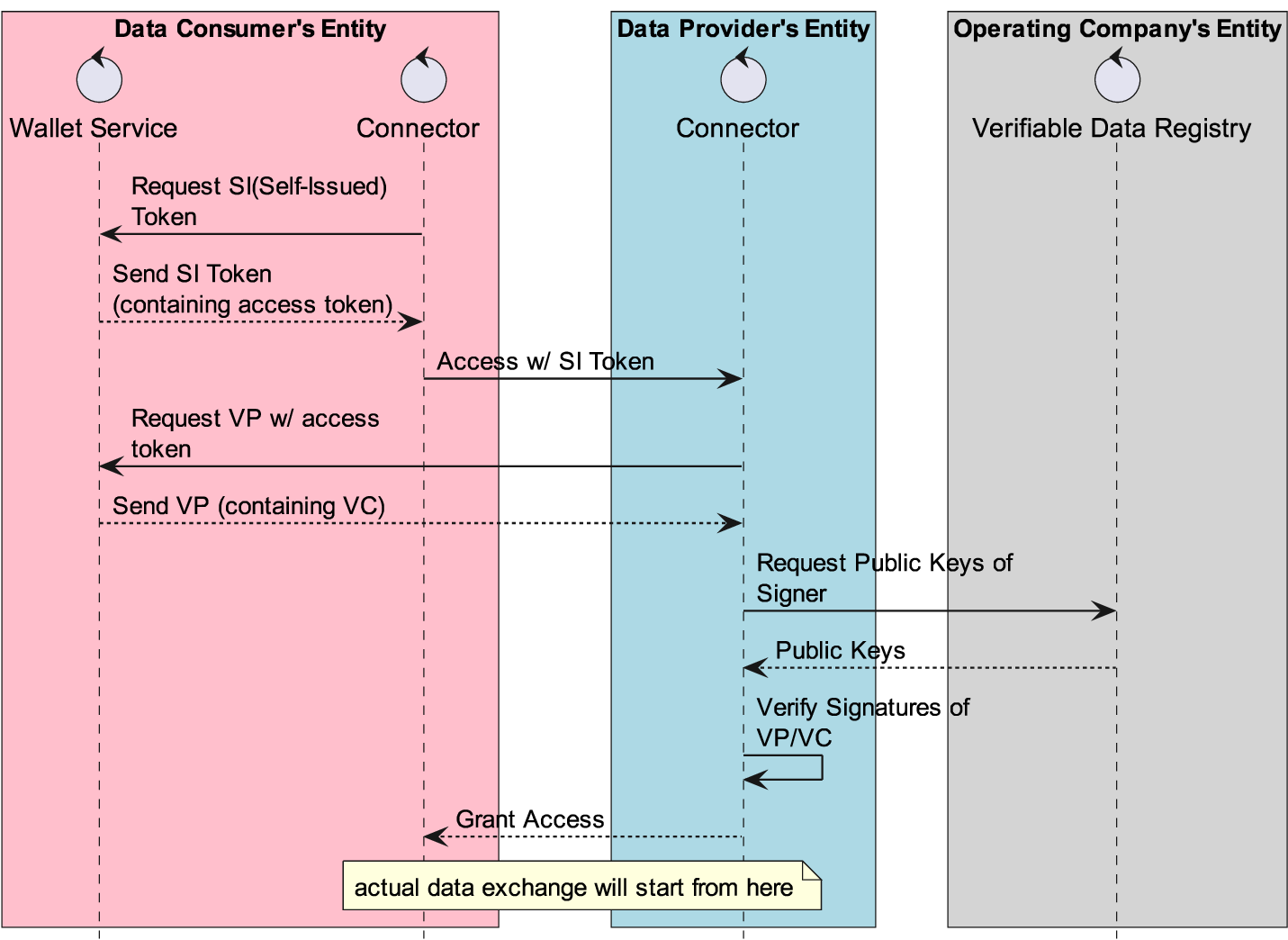}}
\caption{Verification process of the participant in Catena-X.}
\label{fig3}
\end{figure}

On OIDC protocol, the verification is performed by the authorisation server managed by the operating company. In Catena-X, on the other hand, the credential is handled via the Decentralized Claims Protocol (DCP) \cite{DCP}, shown in Figure~\ref{fig3}.

On DCP, the verification is performed by the data provider, who is a verifier. The operating company only provides an issuer's public key through the VDR. The role of the operating company in the verification is a major difference.

\subsection{Device}

As a first step on p1, both platforms validate the device against platform specific standards and register device's endpoint into a connector endpoint resolver. The connector endpoint resolver provides the connector endpoint to the data consumer according to given search keys (e.g. identifiers representing the data provider or its connector). The available search keys are different on each platform, but the endpoint provided by the resolver has assurance that has been validated by the operating company.

As a second step on p1, external Certification Authority (CA) validates the device against common standards (e.g., Extended Validation) and issues a SSL certificate.  This SSL certificate includes domain information used by the endpoint, so it can prove ownership of the endpoint.

In terms of p2, consistency between the endpoint and the device showed by the endpoint can be verified by an SSL certificate.

\subsection{Dataset}

The trust of the dataset depends on the data provider in principle because the dataset is generated by the data provider. Thus, in terms of p1, both platforms do not have any validation function or method but supporting function to improve portability of the dataset across participants. Portability of the dataset here means that one participant can use the dataset from another participant without any conversion. For ensuring portability of the dataset, all participants must adopt same semantics on their dataset. In order to standardise semantics of the dataset in a data space, both platforms have a repository (called as vocabulary repository in DATA-EX, semantic hub in Catena-X) of data models and vocabularies. This repository manages the vocabularies and models expressing datasets in the data space. The data provider organization can use this repository as a reference or propose new vocabularies and models if existing data models do not fit the purpose. On Catena-X also has over 100 pre-defined models and vocabularies with external reference to ECLASS\footnote{\url{https://eclass.eu/en/}}.

In terms of p2, only DATA-EX supports a signable format of datasets, called as Data Distribution Package (DDP). DDP allows the data consumer to verify the creator's authenticity of the dataset.

\subsection{Data catalog}

In terms of p1, for the same reason as the dataset, the trust of the data catalog depends on the data provider and both platforms have a function to keep the data catalogs interoperable in a data space. Both platforms basically restrict the catalog to the DCAT Version 2 \cite{DCAT} format by this function (data catalog creation tool do in DATA-EX, connector do in Catena-X).

In terms of p2, both platforms have no functions to support.

\subsection{Contract}

The trust in the contract should be ensured by both data consumers and providers, as the contract is an object created by both participants. In terms of p1, Catena-X supports the communication between both participants through the connector's contract negotiation API, which is defined in the dataspace protocol \cite{DSP}. Using this API, the data provider can propose a contract candidate and the data consumer can send the confirmation result (agree or disagree). DATA-EX doesn't have this functionality.

In terms of p2, however, the contract should be concluded in a commonly used manner because the contract has legal requirements. For this reason, both platforms recommend concluding the contract outside of the platform (e.g., via an external contract brokering service such as DocuSign). The connectors on both platforms can work with the external contract brokering service.

\subsection{Sending \& Receiving Log}

\begin{figure}[htbp]
\centerline{\includegraphics[scale=0.3]{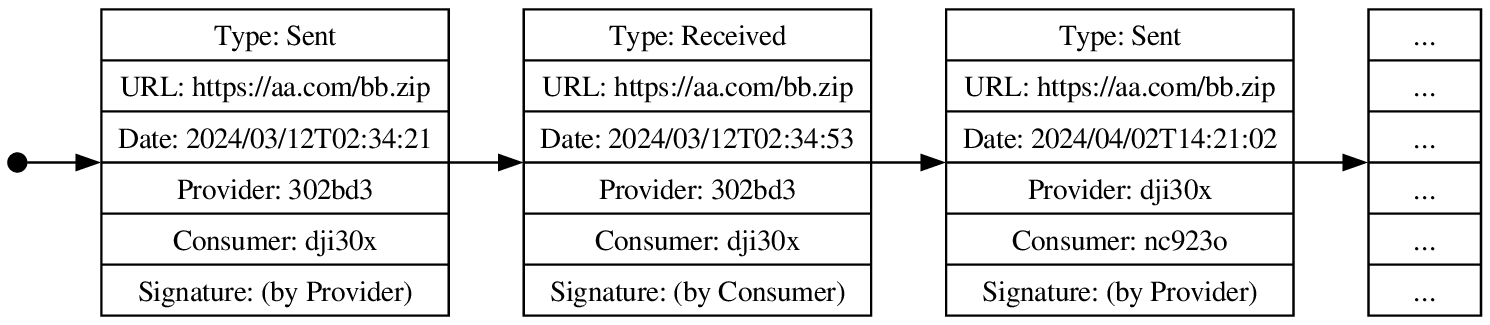}}
\caption{An example of provenance.}
\label{fig4}
\end{figure}

On both platforms, the sending and receiving log is basically generated by the connector and managed in the participant's device. Only DATA-EX has an optional centralised management service (Provenance Management Service). This service stores the sending and receiving logs to construct the proof of origin, which has the structures shown in Figure~\ref{fig4}.

The provenance consists of a repetition of a sent and received log. In terms of p1, by holding digital signatures by both the data provider and consumer in logs, the consistency of the sent and received logs can ensure the occurrence of the data transfer. In terms of p2, the provenance is verifiable as each log is signed by the creator of the log.

\section{\uppercase{DISCUSSION}}
\label{sec:discussion}

Based on the analysis results above, we discuss the challenges of the international connectivity of data spaces. First, we review the gaps identified as shown in Table~\ref{table4}. Second, we discuss an inter-exchangeable topology based on the review. Third, we discuss two challenges based on the topology. Finally, we discuss possible policy alignment for addressing two challenges.

\subsection{Review of Gaps}

\subsubsection{Participant}

It is difficult to close the gap because of the different ways in which trust is ensured and verified. The trust of participants on each platform is not the same, because participants are validated with platform specific standards at generation time. In addition, the difference of verification method makes it difficult to standardise data exchange protocols. Catena-X cannot adopt DATA-EX's OIDC protocol as verification method because Self Sovereign Identity (SSI) is an essential requirement in Europe. Similarly, adopting Catena-X's DCP as a verification method on DATA-EX could take a long time because there are no laws or government guidelines such as eIDAS2.0 on digital identity wallet in Japan.

\subsubsection{Device}

In both platforms, the device is validated in two steps. The first step is the validation at the time of registration with the endpoint resolver. The second step is the validation performed by CA at the time of getting device's SSL certificate.  In the second step, the device is validated with common standards (e.g., Extended Validation) by CA, so certain compatibility will be ensured in trust of the device.

\subsubsection{Dataset}

Semantics of the dataset are different on each platform, so it may spoil portability of the dataset across multiple data spaces. Except for this portability issue, however, exchange of the dataset itself is possible because each platform does not limit semantics of an acceptable dataset to certain one.

\subsubsection{Data Catalog}

Both platforms adopt the same format as a basis and restrict an acceptable catalog format. Principal properties of the catalog are common in both platforms, so it is interoperable to a certain extent.

\subsubsection{Contract}

Both platforms recommend concluding the contract outside of the platform, so trust of the contract is not scope of assurance by platforms.

\subsubsection{Sending \& Receiving Log}

Only DATA-EX assures trust of the sending and receiving log by enabling these logs verifiable in centralised management service.  Verifiable sending and receiving logs can be also used to proof transferred path of the dataset across three or more participants and this may contribute to ensure trust of the dataset.  However, this gap does not disturb data exchange because use of centralised management service is not mandatory.

\subsection{Inter-exchangeable Topology}

\begin{figure}[htbp]
\centerline{\includegraphics[scale=0.32]{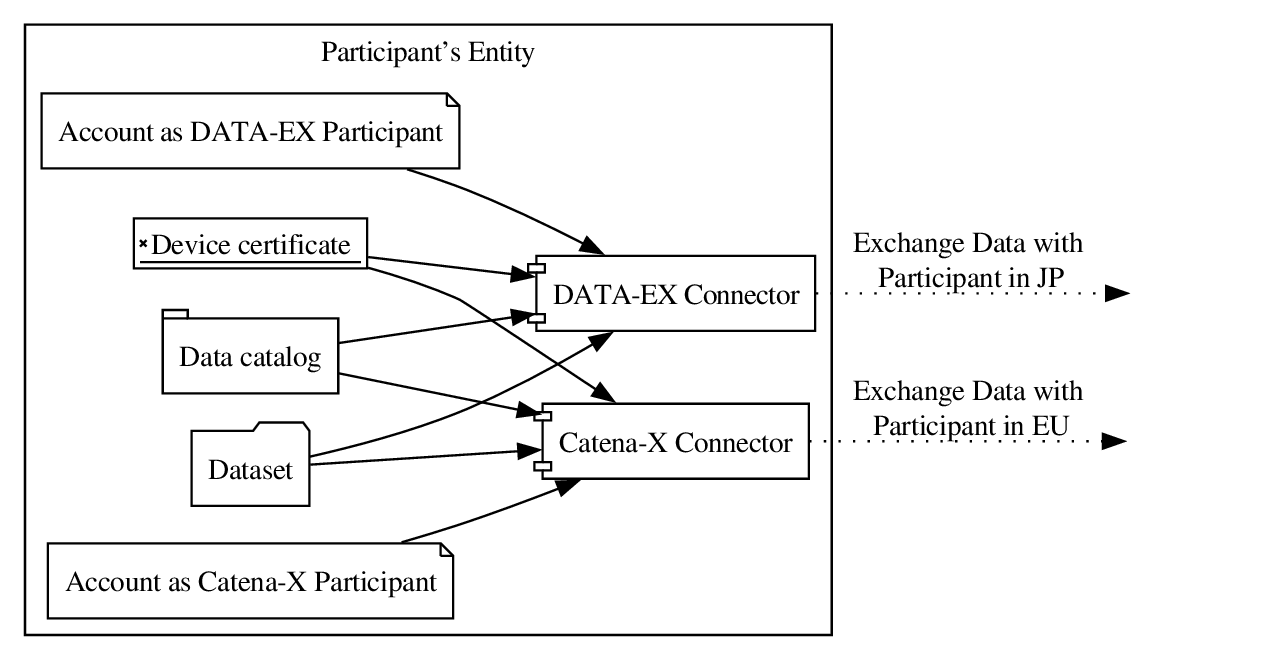}}
\caption{Inter-exchangeable topology between DATA-EX and Catena-X.}
\label{fig5}
\end{figure}

Figure~\ref{fig5} shows the possible inter-exchangeable topology between DATA-EX and Catena-X based on our gap analysis.

To discuss inter-exchangeable topology, we judged each gap whether it interferes interoperability or not. For an object which has the former gap, we thought that we should use the function or method on each platform as they are. In our judgement, only the participant has the former gap, so a connector, which is a component to verify the participant, and an account, which acts as a participant's credential, should be prepared for both platforms like a dual-stack interface. Fast and economic way to prepare the connector and account will be desired because there is large burden in preparation against two platforms.

For the rest objects, we thought that we may make it to be common or omit it. In our judgement, the dataset, data catalog, and the device certificate can be used commonly in both connectors with tiny conversion. The contract and logs can be omitted because these are managed outside of data spaces or not mandatory.

\subsection{A Challenge on The Participant}

\begin{figure*}[htbp]
\centerline{\includegraphics[scale=0.32]{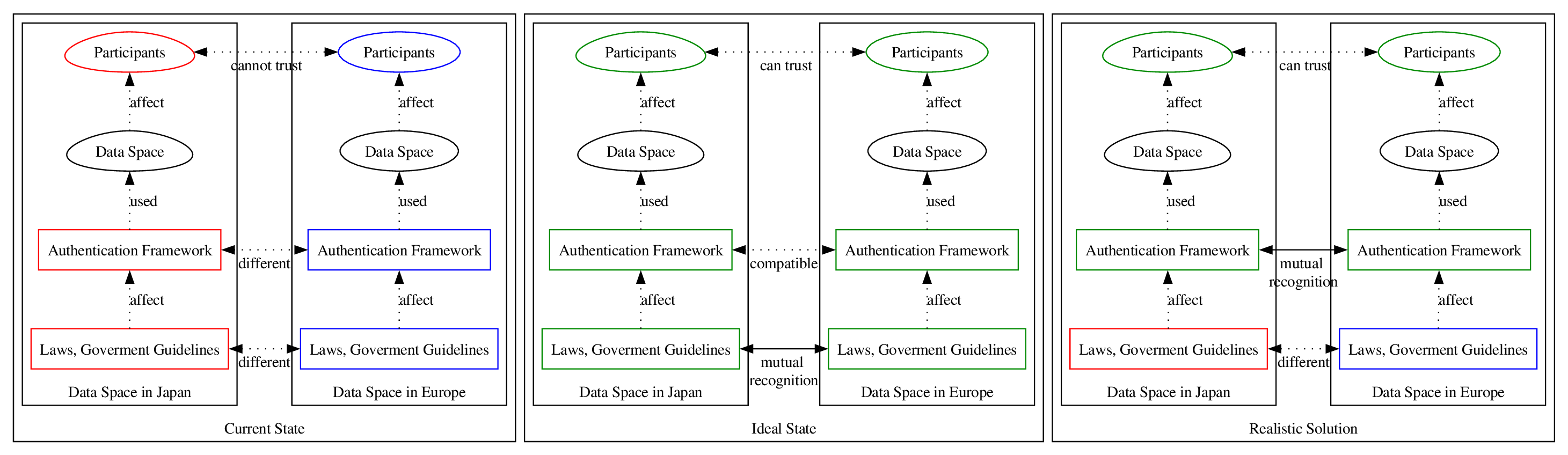}}
\caption{States of a challenge on the participant.}
\label{fig6}
\end{figure*}

Enabling interoperability of the participant trust between data spaces is a high priority challenge, as the trust of other objects depends on the participant trust. Figure~\ref{fig6} shows the current state (left), the ideal state (center), and a realistic solution (right) in terms of the participant trust.

On the current state, laws or government guidelines about digital trust are different in Japan and Europe. Especially, there is no comprehensive one like eIDAS2.0 in Japan. This difference is reflected in the difference of the authentication framework. The authentication framework is used by the operating functions of the data space, so the trust of the registered participants will be different as a result.

In the ideal state, laws and government policies are mutually recognised. Mutual recognition of public (i.e., governmental) leads to compatibility of the authentication framework. As a result, the trust of registered participants will be compatible in some extent. However, this model is difficult to implement at an early stage, as it could take a long time establish mutual recognition. In fact, even electronic authentication and electronic signature, which has been already used in various systems, is not mutually recognised completely among each country \cite{keio}.

Figure~\ref{fig6} (right) shows our proposal as a realistic solution. We recommend that the mutual recognition of private authentication frameworks should precede the mutual recognition of public to achieve early adoption. While this approach needs to continuously adapt to the latest developments in legislation or government policy, it is advantageous for participants who wish to communicate to other participants beyond a country or region.

\subsection{A Challenge on The Dataset}

Another challenge is the semantics of the dataset beyond the single data space. Although the trust of the dataset depends on largely the participant, as mentioned in a previous section, the data space applies governance to the data models and vocabularies by providing the repository and pre-defined models. The repository does not cover a data space in other countries, and the pre-defined models are not mutually recognised between countries. This situation makes it difficult for the participant to trust the dataset in another data space.

To address this problem, we consider two solutions. The first solution is to create an international repository. Although it is an ideal solution, it takes a long time to proceed mutual recognition between countries or regions. The second solution is to create an index repository to show correspondence of semantics among existing repository in each data space.  It is more realistic than the first solution because registration and maintenance of the index can be led by the participant. To implement index repository, some existing semantic technologies such as knowledge graphs \cite{theissen2023semantics} may be useful. Automation of registration or maintenance using technologies such as a name identification method is also suitable for this solution. Thus, we believe the second solution should be proceeded ahead of the first one.

\subsection{Possible Policy Alignment}

For address two challenges above in an ideal way, policy alignment and mutual recognition between countries or regions is essential. There are two approaches to achieving it. The first is standardization. While the IDSA standardise the data space itself and related technologies in the data space protocol, Gaia-X standardise trust required to join the data space ecosystem in the Gaia-X compliance document \cite{gxcompliance}. The Gaia-X compliance document defines trust standards to be satisfied as some levels and laws or regulations to be followed to be certified at each level. Gaia-X itself is a European organization, so approach to develop trust standards applicable internationally through international and neutral standardization organizations is necessary.

The second approach is discussion in the international committee. At G7 Hiroshima 2023, the establishment of the Institutional Arrangement for Partnership (IAP) \footnote{\url{https://www.digital.go.jp/en/policies/dfft/dfft-iap}} was endorsed by G7 leaders to  operationalise DFFT. The IAP can provide opportunities for discussion utsing existing committees or organs of international organisations, so discussing challenges here can be another approach.

\section{\uppercase{Related Works}}
\label{sec:relatedworks}

A basic survey against data space connectors was carried out \cite{dam2023survey} \cite{bacco2024data}. A total of eight connectors including EDC used in Catena-X were surveyed and evaluated, but Japanese connectors are not included in the surveyed connectors. Moreover, it didn't discuss the interoperability issues between connectors, while they explains IDS RAM \cite{IDSRAM4} as a common specification for data spaces. Our work is to compare European connectors in accordance with IDS RAM and Japanese connectors not in accordance with IDS RAM and to consider interoperability among them.

Attempts to create design options for data spaces by analysis of published papers or interview were conducted \cite{giess2023design} \cite{giess2025discovering}. This paper also conducted a frequency analysis of design options against existing data spaces. Their design options include a perspective of trust. They emphasise the dataset and log as objects to consider trust, however, many other objects, such as contracts, are generated in an actual data exchange process. Our work is to conduct an analysis with comprehensive criteria covering a whole data exchange process.

Consideration of interoperability between data spaces was carried out \cite{schinke2024enabling}. They investigated and analysed various technologies (including the data space) that can be used to increase data sharing.  They proposed six ways to enable trustworthy data sharing using the technologies they researched. One of six ways included an architecture that combines multiple data spaces. They stated that handling of a dataset for multiple data spaces is a challenge in this architecture and this challenge should be solved by data trustee \cite{schinke2023trustful} \cite{steinert2024design}. The data trustee can hide a layer of connectors from the participants by creating intermediary layer, so the participant does not have to consider interoperability issues. While our work is to discuss an inter-exchangeable topology without any trust intermediary like the data trustee, the data trustee can be another solution towards achievement of interoperability.

Another consideration of interoperability between data spaces was carried out \cite{hutterer2024scopes}. They surveyed and analysed the scopes of governance in data spaces from published papers. As a result of the analysis, three types of governance were identified: ecosystem governance, technological governance, and operational governance. In the technological governance, the authors argues that the federated data space architecture \cite{schleimer2023architecture} can achieve interoperability across different data spaces. The federated data space architecture expresses the architecture to realise global data availability across multiple data spaces as three logical layers, local foundation layer, federation layer, and global presentation layer. This architecture was evaluated with a circular economy example, but designs for each layer were abstract because no actual connectors were described. Our work is to discuss an inter-exchangeable topology at a more concreate level by comparing and analysing actual connectors.

An experimental environment (called as International Testbed for Dataspace Technology, or ITDT) to try interoperability measures was proposed \cite{matsunaga2023itdt}. The ITDT aims to enable researchers or developers to test connectivity or interoperability among connectors which built on different technological bases. The ITDT can be an adequate environment for us to implement and test our proposals in future, although a DATA-EX connector has not provided yet in the ITDT.

\section{\uppercase{Conclusion}}
\label{sec:conclusion}

In this paper, we analysed DATA-EX and Catena-X as a representative data space platform in Japan and Europe from the perspective of trust. In order to make a fair comparison between two platforms, we decided a common data exchange process and identified six objects that are generated in the decided process and created analysis criteria over the identified objects. Using the analysis criteria, we analysed two platforms and identified gaps on all objects.

Then, we reviewed the identified gaps and concluded that the trust of participant to be the most significant challenge. Based on the review, we discussed inter-exchangeable topology between two data spaces and two challenges. As the first challenge, we identified compatibility of participant trust across data spaces and a realistic solution to it. As a second challenge, we stated the semantics of the dataset beyond single data space and a brief solution of it.

Our study is, naturally, subject to several limitations that must be considered. Because of the analysis method focusing on trust, our solutions do not cover all aspect of interoperability but focus on compatibility of trust on participating entities and the data exchanged. Especially, issues like how to assure compatibility between different connectors is not discussed. Continuous discussion is needed for it considering each standardization activities, although multi-stack topology we proposed can be one solution for a while.

Ideally, both two challenges we stated should be addressed by each other's government and international standardisation organisation, but a work of them takes time to reach consensus. Therefore, in the perspective of immediacy, we believe that private trust initiatives are just as important as public initiatives. We hope this paper contributes to further efforts towards a sustainable and interconnected data economy in the future.



\bibliographystyle{apalike}
{\small
\bibliography{main}}



\end{document}